# Recent advances in stimuli-responsive core-shell microgel particles: synthesis, characterization, and applications

**Julian Oberdisse[1] and Thomas Hellweg[2]**

*[1] Laboratoire Charles Coulomb (L2C), University of Montpellier, CNRS, 34095 Montpellier, France*

*[2] Department of Physical and Biophysical Chemistry, Bielefeld University, Universitätsstr. 25, 33615 Bielefeld, Germany*

*This manuscript is dedicated to Matthias Ballauff on the occasion of his transition to the state of a very active and busy senior-professor in 2019.*

**Abstract**

Inspired by the path followed by Matthias Ballauff over the past 20 years, the development of thermosensitive core-shell microgel structures is reviewed. Different chemical approaches, from hard nanoparticle cores to double stimuli-responsive microgels have been devised and successfully implemented by many different groups. Some of the rich variety of these systems is presented, as well as some recent progress in structural analysis of such microstructures by small-angle scattering of neutrons or X-rays, including modeling approaches. In the last part, again following early work by the group of Matthias Ballauff, applications with particular emphasis on incorporation of catalytic nanoparticles inside core-shell structures – stabilizing the nanoparticles and granting external control over activity – will be discussed, as well as core-shell microgels at interfaces.

## 1. Introduction

Since their first synthesis in 1986 by Pelton and Chibante [1] so-called smart microgels are subject of a steadily increasing number of studies, and in the period from 2018 to 2019 only on the topic of *N*-isopropylacrylamide (NIPAM), which is still the most studied responsive polymer, about 1600 publications can be found in the web of science. A lot of these publications deal with smart microgels. Microgels are fascinating materials because they are able to respond to changes of external parameters like temperature, pH, ionic strength, or external fields by changing their state of swelling [2, 3, 4]. This makes them interesting for applications in sensors [5], optics, and colloidal crystals [6, 7, 8]. Moreover, they are also potentially useful in drug delivery [9], as smart surface coatings, actuators in biomedical devices [10], stabilizers in switchable Pickering emulsions [11, 12, 13, 14], nanoreactors, or smart carriers for catalytically active nanoparticles [15, 16].

Besides NIPAM, other acrylamide monomers can be used to make microgels which also exhibit a so-called volume phase transition temperature (VPTT). The most prominent ones are *N*-vinylcaprolactam (VCL), *N*-isopropylmethacrylamide (NIPMAM), and *N*,*n*-propylacrylamide (NNPAM) [17, 18]. However, also some other monomers appear in the literature from time to time. The main difference between the respective microgels is their different VPTT. More details about copolymer microgels can be found in a review by one of us [19]. Copolymerization of these monomers can be used to systematically vary the transition temperature. Moreover, copolymerization with acrylic or other organic acids can be used to tune and enhance the properties of these colloidal gels [20, 21, 22]. In this area, Matthias Ballauff's group started to contribute at the end of the 90s of the last century [23]. At about the same time Matthias' group also started to work on core-shell microgels [24]. Increasingly complex microgel architectures like core-shell structures are of growing relevance, since they allow to create improved properties and additional functions like e.g. response to several stimuli, or a particular swelling behavior. Starting from pioneering works by M. Ballauff et al. [25, 26, 27], this approach gave rise to increased interest in making nanoparticle-microgel (NP-MG) hybrids for different applications in optics [28, 29], and catalysis [27]. At about the same time at the beginning of the century, doubly thermosensitive core-shell microgels (MG-MG) have also been developed [30, 31]. Both NP-MG and MG-MG particles will be reviewed in this article, with a particular emphasis on their structural properties as seen by small-angle scattering.

A recent review has reported on the evolution of the last three years, in the field of microgel synthesis, topology, biological applications, drug delivery, microgel packing, and theoretical approaches [4]. Another review focusing specifically on core-shell microgels dates back to 2013 [32], while Suzuki and colleagues have proposed a focus review of their many contributions, including heterogeneous structures like Janus microgels and microgel nanocomposites, which may have a core-shell structure and will be further discussed below [33]. Here we adopt a different point of view, starting with the pioneering contributions at the beginning of the century, more or less following the guidance of M. Ballauff's work. We then propose a special focus on recent progress in understanding polymer-based core-shell structures stemming from scattering methods. The article is thus organized as follows. In section 2, we summarize different approaches to system developments of core-shell and multi-domain structures, with various organic or inorganic nanoparticles, up to pure polymer core-shell microgels. The 3rd section is devoted to novel insights into the structure and properties of core-shell microgels, with the different scattering models developed to describe isotropic but non homogeneous structures. This is followed by section 4 dealing with potential applications, in particular catalysis with embedded nanoparticles, of controllable activity via the swelling state of the microgel. Section 5 deals briefly with core-shell microgels at interfaces. At the end, we will conclude and give perspectives for possible future developments.

## 2. The route to stimuli-responsive multi-domain structures

One of the early pioneers in this field was Matthias Ballauff. In absence of photoinitiators, the seeded precipitation polymerization of NIPAM on a non-thermosensitive nanoparticle (NP) surface allowed growing and crosslinking the NIPAM chains, with however imperfections in the grafting of the microgel on the NP [24, 34]. He and his colleagues then transferred a well-established method for the photochemical synthesis of spherical polyelectrolyte brushes to the synthesis of core-shell microgels predominantly with a polystyrene (PS) core and a responsive and crosslinked poly(NIPAM) shell [26]. This method guaranteed grafting from the core particle. The resulting dense PS-core is highly visible in electron microscopy, and the authors could see the temperature-dependent thickness of the grafted p(NIPAM) layer both directly in the TEM-pictures, and its effect via the resulting repulsion between PS-cores [35]. In an earlier article, the scattering model describing such core-shell structures, based on a homogeneous core surrounded by a homogeneous shell, with additional terms for network fluctuations and heterogeneities, has been used to compare the thermal swelling properties of such core-shell particles to the ones of macrogels [24]. This model system continuously attracted interest of scientists in the field, including variations of the chemical protocol determining the type of core-shell morphology [36, 37, 38].

While M. Ballauff's group continued exploiting the particular PS-core/microgel-shell system as tunable carrier for catalytic applications as discussed in section 4, many other nanoparticle-core/thermosensitive shell systems have been created since then. The common approach is usually to introduce a surface-modification of the NPs, typically by hydrophobizing silane coupling agents like methacryloxypropyl-trimethoxysilane (MPS) offering the possibility to start the polymerization of NIPAM or similar monomers from the surface [39]. This technique is also suitable for silica or silica-coated metallic nanoparticles [39]. For example, Nun et al. studied the hydrodynamic radius of a silica-core p(NIPAM) shell particle in a recent article [40], with a focus on the crosslinking density. Their surprising finding is that the size evolution of the NP hybrid with temperature may be non monotonic, presumably due to anisotropic heterogeneities in crosslinking. A similar silane-grafting technology has been used to formulate temperature-sensitive carbon-core/microgel shell particles [41]. When direct polymer-grafting from the particles was not feasible, in some cases silica layers were formed first, followed by silane surface modification and polymerization. This has been the case, e.g., for magnetic nanoparticles. Hematite [42] or maghemite [43] particles have been covered by silica, the surface of which was then used to grow the crosslinked p(NIPAM) layer. Due to the anisotropy of these particles, the alignment of such systems has been studied in external fields [42]. Moreover, the aspect ratio of the particles is then found to be temperature-dependent, due to the isotropic swelling of the p(NIPAM) layer. The same pathway of grafting on a silica layer was followed with non-NIPAM polymers and other magnetic particles, like e.g. manganite [44]. Back in M. Ballauff's group, a yolk-type encapsulation of gold NPs was generated by first growing a silica layer on gold cores, then polymerizing the crosslinked p(NIPAM) layer, before finally etching the silica layer. This technique thus led to individual gold cores freely "swimming" inside a microgel shell, without being covalently connected to it. The consequence is great colloidal stability of the latter, with tuning of optical and catalytic properties by the microgel swelling [45, 46]. A thorough scattering study of amine surface-modified gold cores with a p(NIPAM) shell was proposed by Dulle et al. [28]. By using a combination of approaches, these authors characterized both the density profile and the internal network structure as a function of different shelling steps increasing the shell thickness. With relevance to the discussion in section 3, it is noteworthy that they have used a deconvolution technique for spherical symmetries of the correlation function determined by indirect Fourier transform.

The technique of removing the silica at high NaOH concentrations was also used to produce hollow microgel particles [47], which have more recently triggered quite some interest in their physical properties, in terms of individual structure, and compression in dense assemblies. Dubbert et al. have reported on a detailed small-angle neutron scattering (SANS) modelling study of the hollow cavity [48]. The void is found to be smaller than the original sacrificial silica core, and it is big for highly crosslinked, stiff shells (but then less thermo-sensitive), whereas it becomes rather small (albeit more thermo-sensitive) in the opposite case. The authors conclude that a compromise has to be sought between both properties. In further studies, they added a second outer layer, and thus reach multilayer particles, either core-shell-shell, or hollow-shell-shell (after etching) [49, 50]. Silica-p(NIPAM)-p(NIPMAM) and void-p(NIPAM)-p(NIPMAM) are thus double thermoresponsive. In the presence of the core particle, swelling is restricted by the grafting via MPS onto the silica. After etching of the core, both shells can adapt more freely to temperature changes, and a clear two-step variation is found. These authors underline the interest of the shell-shell geometry: while the outer shell governs the interaction with other particles and stability, the state of the inner one may influence the material encapsulated in the center of the microgel. The study of the compression of hollow microgel particles embedded in dense suspensions of homogeneous microgels has been proposed by Scotti et al. [51]. Using SANS and contrast matching of the regular microgel particles, they managed to specifically observe the response of the hollow ones to pressure by the neighboring regular particles of same elastic properties.

With respect to NP-MG particles, core-shell microgels consisting of both a microgel core and microgel shell [52] introduce a new functionality to core-shell systems: now the core also undergoes a volume phase transition, and by conveying different thermal properties to the core and the shell, e.g. by introducing acrylic acid monomers, tunable swelling properties have been achieved [53]. The step to combinations of acrylamides in the core and the shell was quickly mastered by Berndt and Richtering, who studied the swelling curves of p(NIPAM) and p(NIPMAM) with LCSTs of 34 and 44 °C, used either as cores or as shells [30]. Of course the difference in transition temperatures triggers new effects, the presence and magnitude of which however depends on the relative thickness of the core and shell, as well as on their strength, i.e. the degree of crosslinking. Berndt et al. have presented a SANS study of these effects in 2006 [31]. One of the peculiarities of such a system was pointed out early by Lyon et al. [54]: the shell and the core exert forces on each other, and unequal volume phase transition temperatures imply the transmission of stresses at the interface, and thus impact of the swelling of the core on the shell and vice-versa. By aiming at a higher difference in swelling temperature combining p(NIPMAM) with p(NNPAM), Zeiser et al. opened the road to an unexpected phenomenon, linear swelling [55]. Indeed, the swelling curves measured through the hydrodynamic radius show a peculiar linear domain, where the radius does not change abruptly, but continuously, over a large temperature range located between the two transition temperatures, of the core and of the shell. As shown in Figure 1a, this behavior is robust: normalizing the hydrodynamic radius in suspension or the ellipsometric thickness when adsorbed to a surface reproduces the same linear swelling at intermediate temperatures (see also section 5). One may note that this is only the case when the transition temperature of the shell is below the one of the core – in the inverse case, a two-step transition is found [30]. Incidentally, this remains true in other solvents, as shown by Lee and Bae for a p(MMA-HEMA) in 1-propanol [56]. For comparison, we have also reproduced in Figure 1b the more classical two-step behavior observed in a system where the VPTT of the shell is above the one of the core.

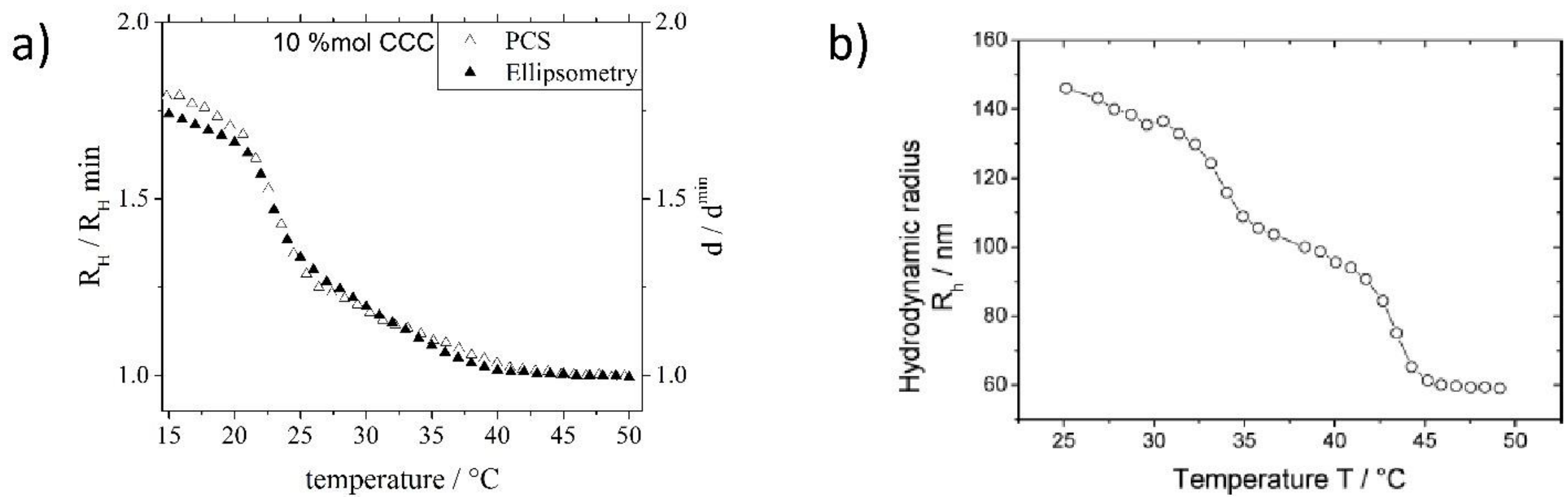

**Figure 1:** Swelling curves of core-shell microgel particles. **(a)** p(NIPMAM) core with p(NNPAM) shell in $H_2O$. Comparison between particles in bulk (PCS) and on surfaces (ellipsometry), normalized to the high-temperature average of either the hydrodynamic radius or the ellipsometric thickness, for core crosslinker contents of 10%mol. **(b)** p(NIPAM) core with p(NIPMAM) shell in $D_2O$. Figure a reprinted with permission from reference [57]. Copyright 2017 American Chemical Society. Figure b reprinted with permission from reference [58]. Copyright 2005 American Chemical Society.

The intriguing linear swelling curves motivated structural studies in our groups [59, 60, 61]. In the framework of the PhD thesis of Marian Cors, the detailed temperature-dependent structure of the core and the shell polymers has been studied by SANS using isotopic labelling. The corresponding model and the main results will be outlined in section 3.

Besides being a force of proposition of general core-shell structures, Matthias' group also pioneered the use of NP-core/MG-shell particles as carriers for nanoparticles, in particular of palladium, platinum, silver, and gold [25, 62, 63, 64, 65]. His group accomplished to show that core-shell particles are able not only to stabilize the catalytically active particles, but also to allow to control their catalytic activity – this will be reviewed in section 4 below. However, it is worth noting in the present section on synthesis that a variety of systems has been designed where nanoparticles are generated within the pre-existing core-shell microgel carrier system serving also as a nanoreactor and template. The microgel is thus created beforehand, and NPs are embedded – on the contrary to the above discussion the NPs do not form the core, and depending on the chemistry, their exact localization may be tuned.

Early work by Suzuki [66, 67] reported on gold NPs being generated in-situ close to the core surface and eventually forming a gold layer of AuNPs using modified p(NIPAM-glycidyl methacrylate) copolymer microgels. These microgels comprise epoxy groups in the core which can be subsequently modified by reaction with 2-aminoethane thiol to introduce SH and also $NH_2$ groups [66, 67]. Brändel et al. optimized catalytic response of silver NPs also located around the core in core-shell microgels with linear thermoresponse the structure of which will be further discussed below [68]. Suzuki and Kawaguchi also achieved high loadings of in-situ synthesized magnetic particles in the spherical microgel core using again copolymerization of glycidyl methacrylate to introduce immobilizing functions [69], whereas Xu et al. provided an example of a cylindrical core-shell structure loaded with magnetite particles stabilized by the shell. They thus gained anisotropy on the mesoscale, which can moreover be oriented with low magnetic fields [70].

By localizing beta-diketone groups capable of complexing and reducing aureate ions in the core of core-shell microgel particles, Thies et al. managed to grow individual gold NPs in each core [71]. Moreover, they used these particles as a seed for further growth, with anisotropic rod-like shapes at high ionic concentrations, and avoiding the formation of secondary gold nuclei.

The tuning of the localization of in-microgel generated nanoparticles has been promoted by Suzuki's group in the past years to design new types of thermoresponsive particles, with thermoresponsive microgels of poly(*N*-isopropylacrylamide) cross-linked with *N,N'*-methylene-bis-acrylamide as cores or seeds for the polymerization which are surrounded by a hard polystyrene shell [72, 73, 74]. Following a seeded emulsion protocol in the presence of PNIPAM microgel particles, the PS particles where found to form raspberry-like structures under high-temperature, surfactant-free polymerization, with however some remaining thermosensitivity, presumably due to incomplete coverage by PS [72]. An example of TEM pictures illustrating such a MG-particle core-shell structure is shown in Figure 2, for different mole percentages of styrene in the feed, and the SDS concentration given in the sample codes. In presence of hydrophobic domains of SDS, the particles lost thermosensitivity. When performing the polymerization of styrene in the swollen state (under appropriate pH-conditions), the PS-particles are found to form a non-compact surface layer. Under pH-induced deswelling, this layer does not become more compact, but NPs seem to diffuse inside the microgel [73]. Finally, in a very recent contribution on the same subject, the Suzuki group presented seeded polymerization of styrene in presence of microgels below the volume phase transition temperature [74]. While raspberry-like structures are again found, the authors point out issues of colloidal stability of the microgels which are related to electrostatic repulsion, which in turn also governs the location of the polystyrene seeds. Last but not least, noteworthy cryo-electron tomography images (and movies) allow understanding of the three-dimensional arrangement of the polystyrene particles inside the microgels.

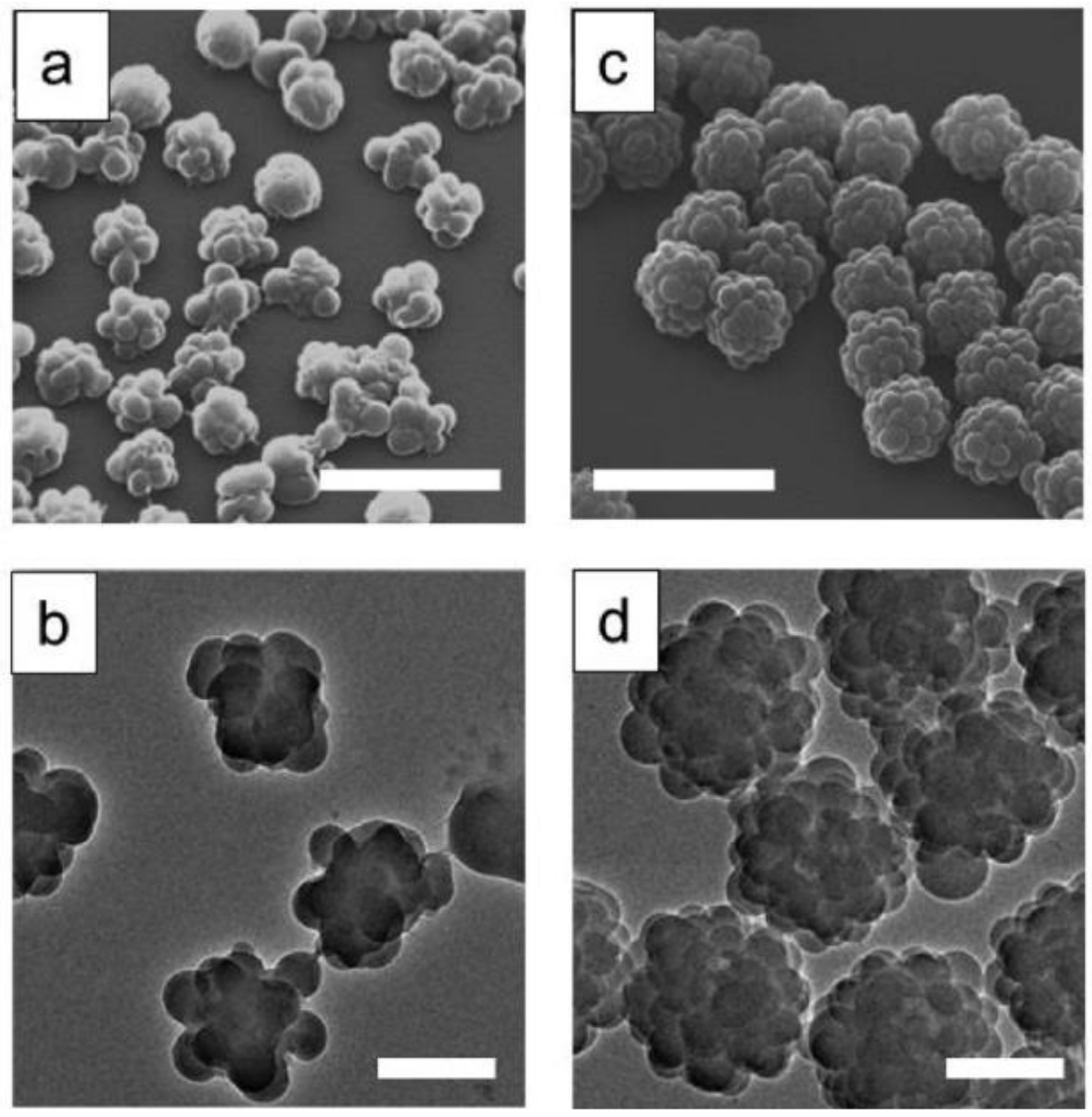

**Figure 2:** Electron micrographs. (a, b) NIPAM-styrene composites NS200(0.5), (c, d) NS200(6.5), and dried on solid substrates at 25 °C. The first number represents the styrene concentration in mmol in the feed, the second one the SDS concentration. (a, c) FE-SEM images. Scale bars are 1 µm. (b, d) TEM images, scale bars represent 300 nm. Reprinted with permission from reference [72]. Copyright 2014 American Chemical Society.

## 3. Internal architecture of core-shell microgels

In a first approximation, most microgel particles are globular, and are conveniently described as spheres. Photon correlation spectroscopy (PCS) gives the hydrodynamic radius of these spheres, whereas small-angle scattering measures the radius of gyration, or the corresponding equivalent hard sphere radius. Comparison of both immediately shows that the microgel particles are in general not homogeneous hard spheres, but possess a temperature-dependent gradient profile, with, e.g., dangling chains at the surface. In scattering, this is immediately visible in the high-q range, where deviations from a Porod $q^{-4}$-law are easily measured. Indeed, a progressive radial decrease of density may be caused by different monomer and crosslinker reactivities: microgels are thus spatially inhomogeneous and possess a fuzzy surface [75], unless crosslinker is added continuously to reach a more even distribution [76]. In the first part of this article, "multi-domain" thermosensitive microgel structures have been reviewed following the original pathway of Matthias Ballauff. Some of them are well described by a core-shell structure, or, more general, a non monotonous density profile. As described in section 2, a particular class of core-shell microgels is obtained by the combination of largely different VPTTs, where the core swells first when decreasing the temperature [55]. The outstanding property of such core-shell topologies [32] is the linear evolution of their radius as a function of temperature between the two polymer phase transition temperatures as shown in Figure 1a. [55, 57, 77]. This linear response may be favourable in particular for sensors and actuators, but it remains unclear how this property is related to the structure of the shell. A possible explanation of the limited core swelling between the two VPTTs has been proposed and has been called the “corset effect” [55]. This concept claims that the shell embedding the core is not swollen at the same temperature, thus restricting the volume phase transition of the core. It is unclear, however, how well this structural picture of a well-defined shell restricting the core is appropriate. It is the aim of the present section to review the corresponding small-angle scattering and modelling approach.

The synthesis of complex architectures of microgels thus triggers the necessity of structural characterization. SANS is well suited, and it is particularly useful for particles suspended in solvent at various temperatures. This technique is capable of simultaneously measuring both average particle sizes and polydispersity, as well as the local structure of macromolecules. SANS and isotopic substitution may reveal the distribution of monomers within particles, as for example in core-shell microgels having one monomer deuterated, which is the ultimate goal of our approach extracting radial density profiles based on a form-free model. In this section, we will discuss modelling approaches to small-angle scattering from such systems. Contrast matching through the solvent and employing isotopic substitution of either shell or core monomers enables studying selectively only one monomer in SANS, and thereby independent measurements of their profiles. One should note that deuteration causes small differences in swelling, as published by us for p(NIPMAM) [60], but its impact remains negligible far from the VPTT. We add that although small-angle scattering is not the only technique to observe microgels [78, 79, 80], it is indeed very powerful due to the possibilities of deuteration allowing differentiating core and shell polymer, in bulk suspensions, without fluorescence labelling.

### 3.1 Core-shell models for small-angle scattering

The scattering of isotropic particles is usually described as a sum of terms, the first of which is linked to the density profile, while the others account for chain scattering, as well as for heterogeneities and dynamic fluctuations [81]. Usually, a Lorentzian term is used to describe the internal polymer mesh, with a mesh size extracted from fitting the correlation length, but other polymer scattering terms like Debye functions or generalized Gaussian coils may be applied empirically [59]. Single compartment microgels (“core-only”) can be modelled in a first approximation by sphere scattering – one of the characteristics of which is the steep high-q Porod law, which can be masked by internal and chain contributions, or by fuzziness of the interface, which also smears oscillations, just like polydispersity. For more complicated geometries, the density profile may be defined piecewise, and constant densities in the core and in the shell are an obvious starting point to describe core-shell particles. Again, introducing fuzziness allows a more realistic description of the interface, and most differences between such fixed-form models actually stem from the type of decay describing the interfaces. Often the decays are described by Gaussians or piecewise polynomials, like parabolic decays. By combining piece-wise parabolic functions, different geometries can be constructed, and namely an empty shell can be described easily [82, 31, 83]. For completeness, let us mention another model based on Flory-Rehner theory recently developed by Boon and Schurtenberger. In p(NIPAM) microgels, this model gave access to the non-homogeneity of the spatial distribution of crosslinker within microgel particles [84].

In the past, such approaches have been applied successfully to describe the structure of core-shell microgel particles investigated by small-angle neutron scattering. The resulting scattering curves were analysed with different core-shell models based on the above-mentioned concepts [30, 58, 31, 82]. In these experiments the absence of isotopic substitution did not allow differentiation of the core and shell monomers. Such measurements and analyses showed the presence of both monomers, and appeared to endorse the intuition that the shell is located just outside the core – we will see below that depending on the system, this is not necessarily the complete picture. The force of these core-shell models is that they are given by analytical formulas of the radial density profiles. It is thus straightforward to Fourier transform them and compare the result with the experimental intensities, and proceed with fitting, needing to fix only a small number of parameters: the location/radii and densities of the core and the shell, and possibly interfacial fuzziness [31, 82, 83]. As already discussed, by adding additional layers, core-shell-shell particles have been synthesized. After silica core etching, hollow-shell-shell structures have been observed[49, 50]. Light scattering has been discussed qualitatively - these authors conjecture that a generalization of core-shell models with different shells might have been able to quantitatively account for the observed intensity features, in particular for the shift in first minima [49]. SANS and detailed modelling has been used to unravel the structure of hollow microgels and their precursors, silica-microgel core-shell particles. By varying the isotopic composition of the solvent between light and heavy water, the contributions of the silica core and of the protonated polymer layer to scattering could be varied via their respective contrast. In particular, Dubbert et al have succeeded in fitting a single model discriminating the inorganic silica core from the shell under all contrast situations [48]. Finally, by comparing this result to the one of hollow microgels, the evolution of the shell with the etching of the core could be followed in detail [48].

The common point of the various fuzzy-sphere models proposed in the literature discussed above is that predefined parameters fix and limit the possibilities for unexpected structures. For example, a given number of shells may be used as a starting point, with possible variation of thickness and position of the interfaces. In some cases, unforeseen fit results may motivate an extension of the model, e.g. by adding a shell. In some occasions, the choice of a given model has been guided by a first form-free modelling [85]. In others, the form-free method proved to be essential to discover an unusual density distribution – microgels less dense in the center – with ultralowly crosslinked particles [86]. "Form-free" modelling is obtained by the construction of any general profile based on the densities in (thin) concentric shells. The intensity can thus be calculated easily by analytical Fourier transformation. Such models have also been applied to copolymer micelles [87]. In Virtanen et al. and also in our approach outlined below, the method has further been extended to include particle polydispersity [86].

### 3.2 Form-free density profiles of spherically symmetric microstructures

In recent contributions, we have implemented a reverse Monte Carlo (RMC) optimization of general, form-free density profiles Φ(r) [59]. Such arbitrary profiles allow the description of SANS intensities without a priori parametrisation. The monomer density profiles evolve stochastically, driven by a χ-square minimization. The aim is to reach agreement between the experimental and theoretical scattering functions. The resulting densities have been demonstrated to be robust with respect to starting conditions or other simulation parameters. In a first article, we have confronted the density profiles of the most simple "core-only" microgels to the models of the literature, and a performance equivalent to the well-established (parabolic) fuzzy-sphere model was evidenced. In this approach, it was supposed that the microgels are polydisperse as measured by AFM, and spherically symmetric. The contribution of polymer chains is visible in the mid- to high-q range, and is successfully described by the generalized coil model [88].

Our aim of understanding the outstanding linear swelling properties of core-shell microgels motivated an in-depth form-free modelling of their internal microstructure. This could be achieved with contrast-variation SANS combined with the determination of the core and shell density profiles. Due to the unexpected geometries, it was important to be able to analyse the data without any a priori assumptions on the shape of the density profiles. Our approach combined several measurements: at first a pure p(NIPMAM) core ("core-only"), onto which either deuterated or hydrogenated p(NNPAM) shells have been grown. It is stressed that all cores are the same in our studies [57, 59, 61, 89], i.e. the second synthesis step of the shell has been performed with pre-existing cores, enabling thus direct comparisons.

In our studies, the monomer density profile Φ(r) is defined on $N_p$ = 30 homogeneous shells of fixed thickness set to ΔR = 2 nm as building units. The profile Φ(r) is then transformed into the scattering length density ρ(r), and ultimately into the scattered intensity I(q). Through the low-angle intensity and particle concentration, the total number of monomers in a microgel particle is known. In some cases, interactions between particles generate a very weak structure factor appearing in the low-q domain, in particular below the VPTT where microgels are swollen. For this reason, usually very low volume fractions are measured; moreover, it has been checked that the results are virtually identical when cutting off the low-q oscillations [61].

The physical model and the elementary reverse Monte Carlo steps of the algorithm are illustrated in Figure 3a. Several different initial conditions in terms of distributions of monomers across the shells have been produced in order to verify the robustness of our approach. The progression of the algorithm is shown in Figure 3b. Monte Carlo steps are performed by random moves of groups of $N_{move}$ monomers from any shell to another, respecting maximum packing of 100%. This quantity is seen to be decreased to one with simulation time, progressively fine-tuning the profile. The deviation between the theoretical curve and experimental intensity is measured by $\chi_i^2$. In the inset, the initial guess and the final fit are compared to the experimental intensity. In order to find a physically acceptable solution to this ill-posed problem, a compromise based on the L-shaped "stability plot" with smoothing term $\chi_r^2$, see [86], was sought by systematically varying the relative importance of fit quality and profile smoothness. These functions are also shown in Figure 3b, $\chi_T^2$ representing the virtual annealing temperature allowing progressive freezing of a solution.  The result of our simulation is a smooth volume fraction profile of monomers leading by Fourier transformation to an intensity prediction compatible with the experimental intensity.

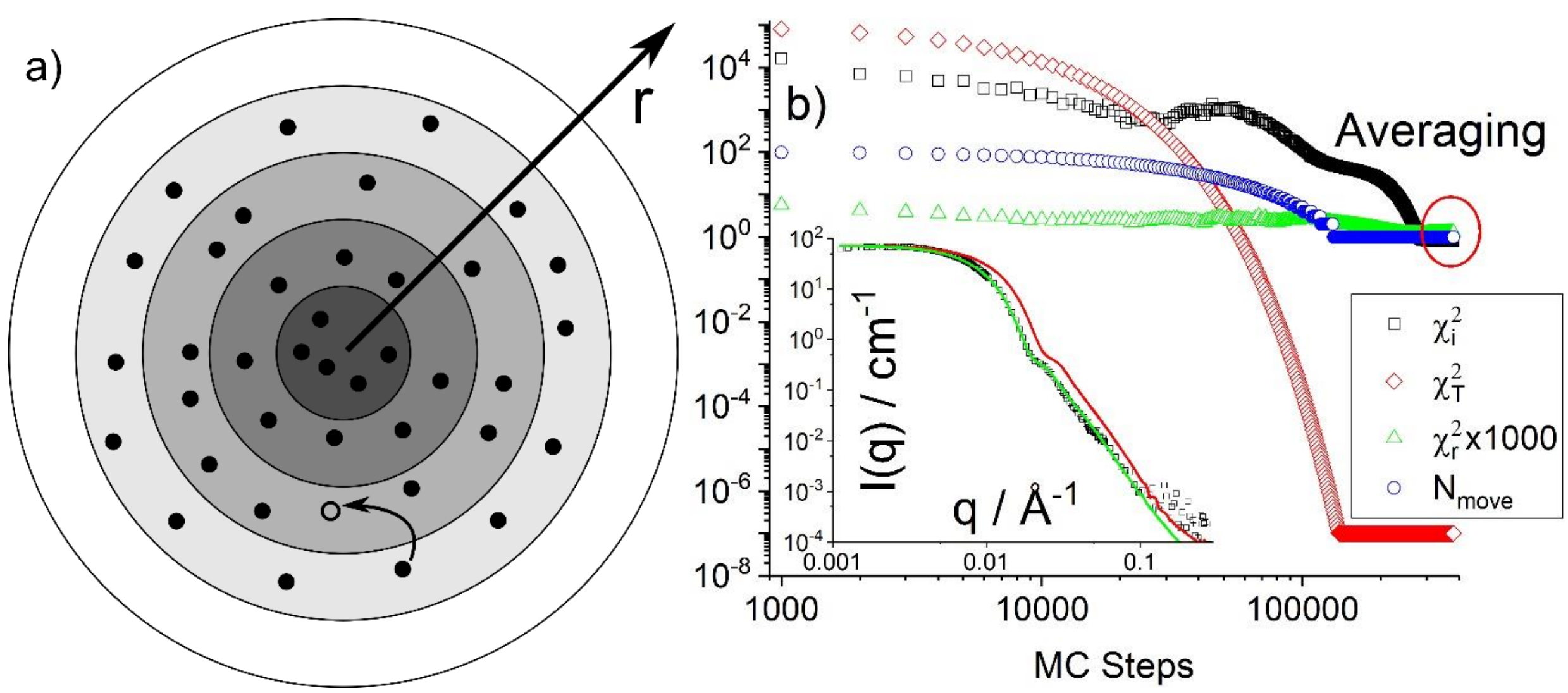

**Figure 3:** **(a)** Shell model of average microgel particle indicating moves of groups of monomers between shells. (**b**) Simulation parameters as a function of the number of MC steps ($\chi_i^2$, $\chi_T^2$, $N_{move}$, the number monomers moved in one step, as indicated in the legend. The lowest value is 1). Once $\chi^2$ falls below 1 the profiles and fits are averaged over the next 100 000 steps. In the inset, the initial guess for the intensity (in red) and the final simulation result (in green)

are compared to the experimental intensity (black). Reprinted with permission from reference [59]. 

We have first investigated the density profiles of the "core-only" system synthesized using NIPMAM with a LCST of about 44 °C [59], and which has been used as the core of the deuterated core-shell particles later [61, 89]. We have first performed PCS and direct imaging providing details on polydispersity and shape which are necessary to set parameters for the SANS analysis. An example of this analysis showing the polymer chain and density profile parts of the scattered intensity is shown in Figure 4a. The resulting profiles describe the "core-only" microgels as a function of crosslinking and temperature (shown in Figure 4b), for subsequent comparison with cores within added shells. This representation of the particle morphology gives direct access both to the internal density (which is directly linked to the amount of water) in the center, and to the details of the particle interface, which is shown to become smoother with decreasing temperature.

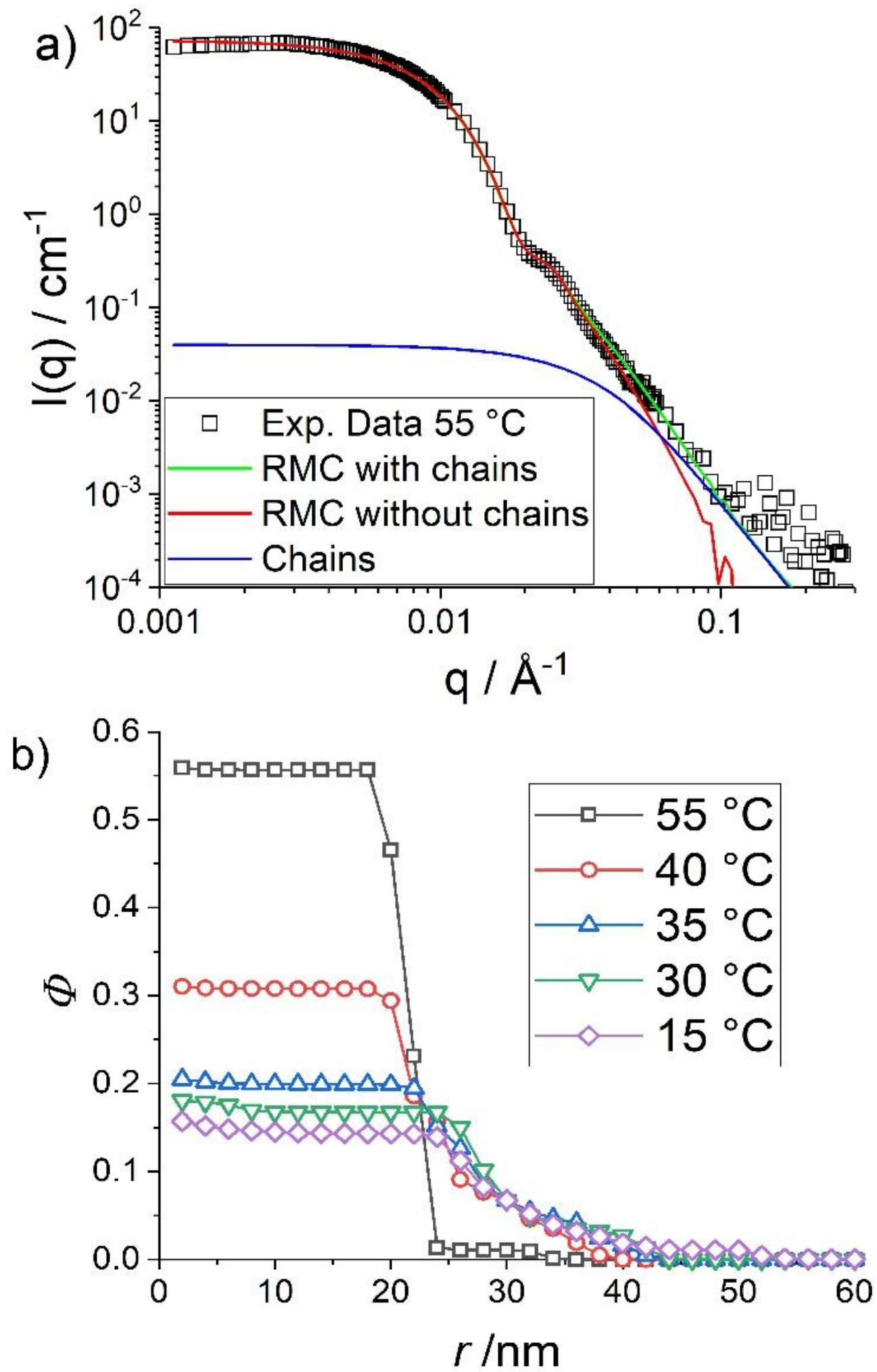

**Figure 4: (a)** Fitted chain and profile contributions to scattered intensity of a microgel (CC = 10 mol%) at 55 °C. **(b)** Density profiles of the H-p(NIPMAM)-core with a CCC of 10 mol% in $D_2O$. Adapted with permission from reference[59]. Copyright 2018 American Chemical Society.

In a following paper, we have reported on the synthesis and scattering analysis of particles made of a p(NIPMAM)-core and a p(NNPAM)-shell [61]. These experiments have been set up to verify the spatial structure of the core embedded in a shell. We have based these measurements on the NIPMAM core monomer used by Berndt et al. (VPTT ≈ 45 °C), [83, 90, 18, 91, 92, 60] but have taken NNPAM monomer (VPTT ≈ 23 °C)[60] for the shell synthesis, thereby generating the linear swelling. Berndt et al. synthesized the shell with p(NIPAM) and its VPTT of ca. 33 °C, [93, 94, 90, 91, 60] thus the linear region is larger in our case, extending from ca. 20 to 45 °C. These microgels are similar to the ones investigated by Zeiser [55], but they are smaller and possess H-D contrast. The smaller size with respect to Zeiser et al. allows measuring the entire form factor in a standard small-angle scattering experiment, and thus gives access to a more detailed analysis of the measured intensities as described below. It becomes thus thinkable that the underlying nature of the corset-effect may be understood on a structural basis. For the H-core/D-shell microgels, measurements in 12 v%/88 v% $H_2O/D_2O$ are of particular interest. They have been designed to contrast-match the deuterated shell polymers and investigate the core structure in presence of the invisible shell chains. Recent data from IR-spectroscopy indicated interpenetration of the core with the shell polymers [77], further motivating a detailed look into the spatial extension of both monomers. It is again underlined that the "core-only" microgel particles investigated in our previous paper are identical to the ones studied here, i.e. the shell synthesis has been performed with the same samples as in [59], and otherwise identical conditions as in the article by Zeiser et al. [55] We have measured a series with various core cross-linker contents (CCC), for several temperature ranging from 15, to 30, 35, 40, and finally 55 °C, encompassing the VPTT of core and shell polymers.

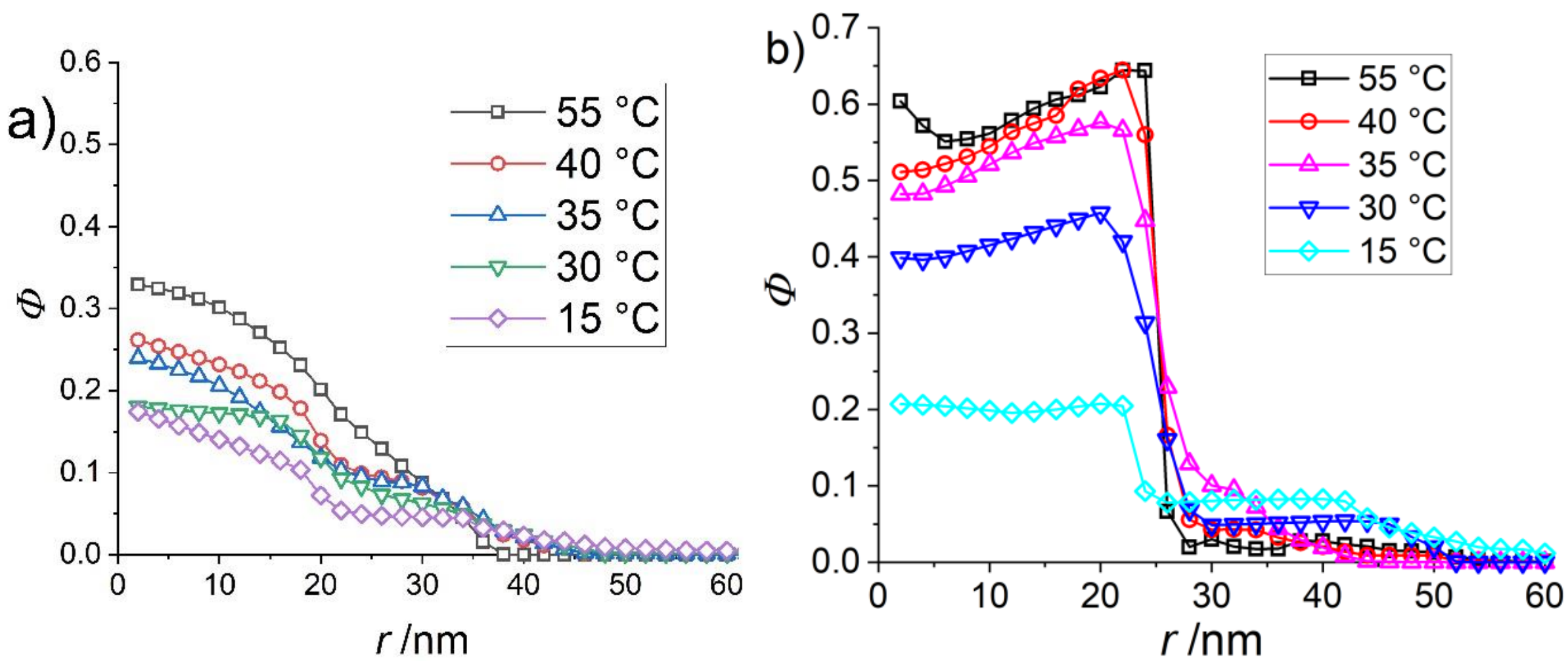

**Figure 5:** Results of SANS analysis (CCC = 10%mol). **(a)** Density profiles of the core monomers of the H-p(NIPMAM)-core D7-p(NNPAM)-shell system with an index-matched shell ($H_2O/D_2O$ = 12v%/88v%). Figure was reproduced from [61]. **(b)** Density profile of the D7-p(NNPAM) shell monomers of same system with contrast matched core ($H_2O/D_2O$ = 77v%/23v%). Fig. 5b was reproduced from [89] with permission from the The Royal Society of Chemistry.

As illustrated in Figure 5a, the unexpected result of our analysis is that the cores embedded in core-shell microgels occupy a larger space than the original "core-only" microgels [61]. Due to conservation of monomer numbers – this is exactly the advantage of using the same cores – the cores possess the same mass and thus the same low-q intensity in absence of interaction. They are thus characterized by a lower volume fraction, which contradicts the prediction of core compression by the (hypothetical) corset effect. This unexpected experimental result could only be rationalized by swelling of the core by the shell monomers.

In the same paper, we have estimated the shell profiles without having the corresponding measurements with core-matching. [61] This was achieved by comparing the H-H with the H-D monomer density profiles, which by subtraction provide the D-profile. Based on the assumption that H-H and D-D syntheses gave rigorously identical particle size distributions, the location of the monomers of the shell and thus their radial density profile could be estimated indirectly. In a third article, we have then directly measured the radial density profile of the shell monomers by contrast-matching the hydrogenated core monomers in 77v%/23v% $H_2O/D_2O$ solvent mixtures. The resulting SANS curves have been analysed by applying our form-free multi-shell reverse Monte Carlo procedure to the data, and the result is shown in Figure 5b [89].

From the comparison of Figure 5a and 5b, which have been obtained on the same microgels for different solvent mixtures (different scattering contrast), it results that the core monomers are diluted in the center and therefore extend further towards the outside. This is caused by the presence of the (so-called) shell monomers, which are seen in Figure 5b to reach deeply into the core. The shell monomers form both a small external shell and an internal insoluble skeleton with a density gradient which restricts the swelling of the core once its VPTT is reached. For comparison, a homogeneous blend of core and shell monomers within the microgel particle would generate a single swelling step, whereas regions of varying monomer ratio would respond at different temperatures. It would be interesting to conceive a mechanical model describing the swelling and shrinking of such heterogeneous microgel particles. Indeed, one may conjecture that it is this gradient structure which causes the linear swelling between the VPTTs of the monomers involved. Moreover, it can be guessed that the linear swelling and shrinking property is intimately related to the exact shape of the density gradient.

## 4. Applications of core-shell microgels in catalysis

The use of core-shell microgels was pioneered by the Ballauff group [25, 27, 95, 62, 96, 15, 16] and still represents an active area of research which is close to potential applications. Especially in catalysis based on metal nanocrystals, core-shell microgels are very useful, since they stabilize the nanoparticles and prevent aggregation. This allows for an easy separation of the hybrid particles from the reaction mixture which enhances largely the reusability of the catalysts and leads to a straightforward work-up of the reaction mixture. This topic was addressed by Suzuki and coworkers [97].

Moreover, as also already shown by Matthias Ballauff and co-workers, the responsive hydrogel part of such hybrid materials can be used to produce non-Arrhenius behavior which is a sign for the control of the catalytic activity [25, 27, 95, 63]. Meanwhile several authors tried to improve the carrier particles, and the recent publications in this context will be presented in the following.

As already mentioned, Suzuki et al. have prepared core-shell poly (NIPAM-co-glycidylmethacrylate) microgels with a rigid core mainly consisting of the glycidyl methacrylate (GMA) comomer [97]. The core-shell structure is obtained in a simple one-pot precipitation polymerization due the faster polymerization of GMA. In some cases, the authors added 2-amino-ethanol to the reaction mixture leading to a localization of positive charges at the interface between GAM core and p(NIPAM) shell. This is afterwards helpful to generate catalytically active nanoparticles (NPs) which are localized at this internal interface. In total five different hybrid systems with Au NPs were made. Also Suzuki et al. use the reduction of *p*-nitrophenol as model reaction to follow the activity of the catalyst. Hence, the results are nicely comparable to the works by Ballauff et al.

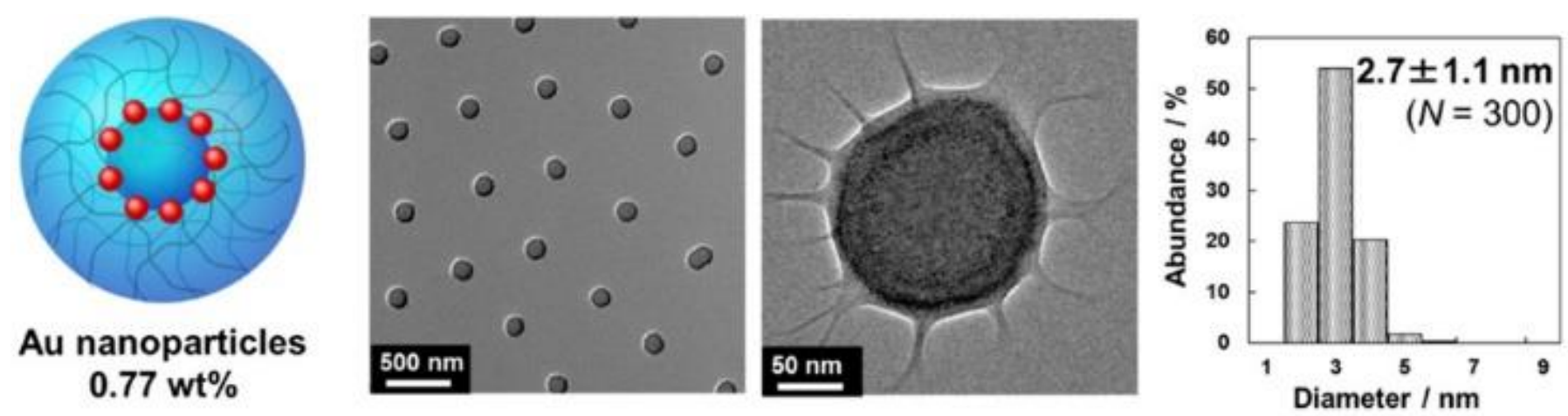

**Figure 6:** Image of the successful system from the work of Suzuki et al. This core-shell system shows excellent re-usability as catalyst. Reproduced from Ref. [97] (Open Access Publication in ACS Omega).

One of these systems, shown in Figure 6, exhibits excellent re-usability and even after the fifth catalytic cycle there is nearly no loss in activity observable. In between the catalytic reactions the hybrids are separated from the reaction mixture using centrifugation steps.

Shuli Dong and co-workers [98] have prepared another core-shell system which can be used for catalysis. Also this system has a hard core. In this case the core consists of silica-coated Au nanorods which are used as seeds in the microgel synthesis. As in some of our old works MPS is used to enhance the incorporation of the inorganic core into the shell [39, 99]. In addition to NIPAM the authors use 1-vinlyimidazole as comonomer. This enhances the binding of spherical gold NPs which the authors generate from $HAuCl_4$ by reduction using $NaBH_4$ in presence of the core-shell microgels. These particles present a hierarchically organized functional structure and the nano-rod core provides the possibility to shrink the particles by irradiation in the near-infrared (see also related works by Kumacheva et al. [100, 101]). Hence, light can be used to externally control the degree of shrinking of the shell of these particles and hence, to control the activity of the spherical Au NPs embedded in the shell. Shuli Dong et al. also use the reduction of *p*-nitrophenol to measure the catalytic activity.

Another article which also uses the same model reaction was published in 2019 by Tzounis and co-workers [102]. In this work the authors describe the formation of a hierarchically ordered core-shell microgel with a core made of silver and gold covered by a cross-linked p(NIPAM-co-allylamine) shell. In this thermoresponsive shell, silver nanoparticles are generated by reduction and prevented from bleeding by the allyamine. Hence, the particles have the following structure: AuAg@p(NIPAM-co-allylamine)@Ag. The distribution of the Ag particles inside the polymer shell is described by the term “satellites” by the authors.

The AuAg cores are obtained by first making a Au@p(NIPAM-co-allylamine) system and subsequently overgrowing the gold cores with silver. In the reduction of *p*-nitrophenol this new structure seems to have superior turnover compared to other microgel nanoparticle hybrids and very high rate constants are achieved.

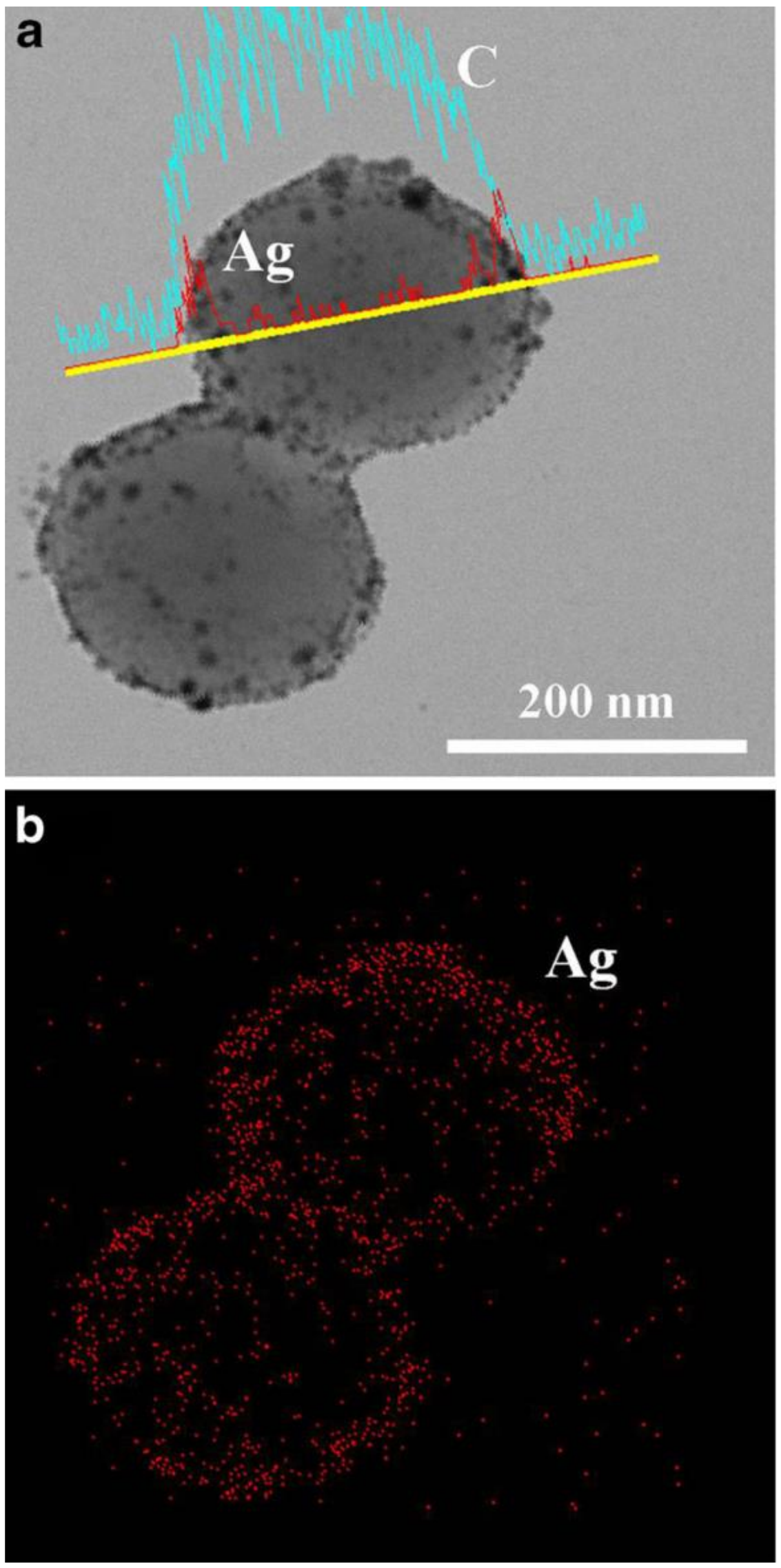

**Figure 7:** High resolution TEM (a) and STEM (b) image of the particles prepared by Yang et al. The STEM-EDS image nicely reveals the distribution of the silver nanoparticles in the hybrid core-shell microgels. Reproduced with permission by Springer Nature from Ref. [103].

Another way to achieve stronger persistence of the nanoparticles in the microgel network is the use of carboxylic acid groups. Based on this idea Yang et al. have prepared core-shell microgels with carboxylic acid groups in the shell [103]. The microgel particles have thus a p(styrene-co-NIPAM) @ p(NIPAM-co-methacrylic acid) structure and are made in a two-step synthesis. The shell has a remarkably high nominal methacrylic acid content of 30 mol% with respect to NIPAM. Following the synthesis, the core-shell microgels were dried and redispersed in ethanol. In this medium the authors reduced $[Ag(NH_3)_2]^+$ to generate the Ag NPs. The obtained hybrid particles were characterized by TEM and Figure 7a shows a typical image. In Figure 7b a STEM-EDS image is shown. Here, the silver particles appear as red dots. The silver particles are mainly formed in the shell indicating that the methacrylic acid has an influence on the Ag localization. Also Yang et al. have used the reduction of *p*-nitrophenol to study the catalytic activity of the new core-shell hybrid microgels and the observed behavior is similar compared to the findings presented in Matthias Ballauff’s pioneering works on this topic. However, the swelling curves of these particles seem to exhibit a two-step transition which is different compared to the PS@p(NIPAM) particles used by Matthias.

Also one of us has recently started to work on catalytically active core-shell microgels [68] using carboxylic acid groups to achieve better fixation of the metal NPs as well. In this work acrylamide-based, thermoresponsive core–shell microgels with a linear phase transition region are exploited as improved carriers for catalytically active silver nanoparticles in this study. The particles are slightly different compared to those described in [57]. The following general microgel structures were used: p(NIPAM-*co*-AAc)@p(NNPAM), p(NIPMAM-*co*-AAc)@p(NNPAM), and p(NNPAM-*co*-AAc)@p(NNPAM). In all cases the core contained acrylic acid as comonomer to improve the binding of the metal NPs. In this context, the swelling behavior of the carriers and the stability of silver NPs inside the polymer network was investigated by photon correlation spectroscopy, transmission electron microscopy, and by following the surface plasmon resonance of the NPs. Depending on the cross-linker content of the microgel core, very good stability of the NPs inside the microgel network was observed, with nearly no bleeding or aggregation of the NPs over several weeks for core cross-linker contents of 5 and 10 mol%. The architecture of the hybrid particles in the swollen state was investigated with cryogenic transmission electron microscopy. The particles exhibit a core–shell structure, with the silver NPs located mainly at the interface between the core and shell. Such a hybrid architecture was not yet used before and leads to improved stability of the Ag nanocrystals inside the microgels. Also in this work reduction of *p*-nitrophenol was used as model reaction. By studying the reaction at different temperatures a good switchability of the catalytic activity was revealed as shown in Figure 8.

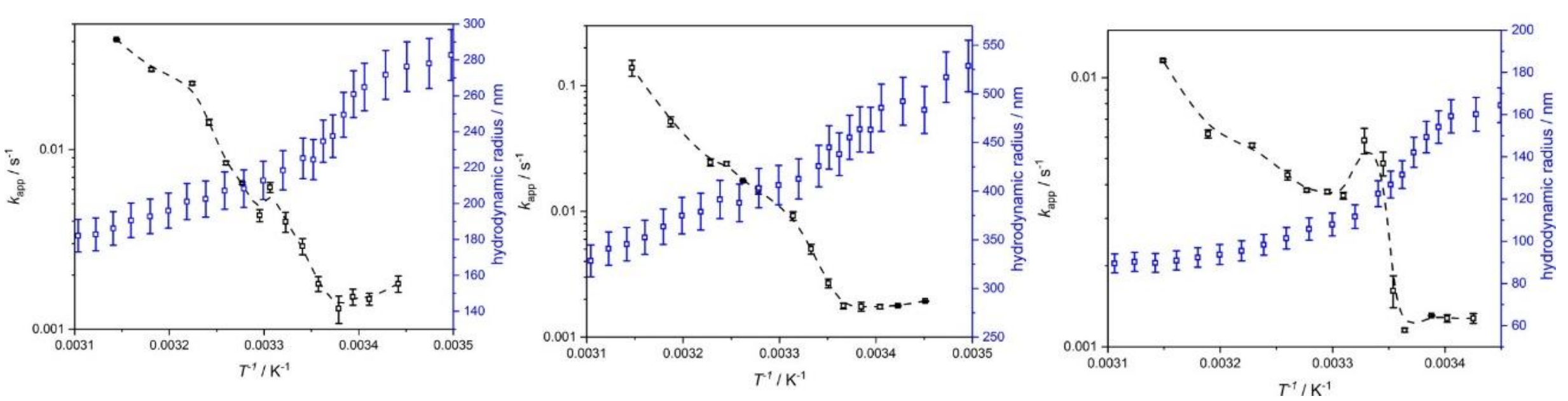

**Figure 8:** Arrhenius plots for the different synthesized microgels showing the enormous influence of the swelling/shrinking state of the core-shell particles on the activity of the Ag NPs. The figure presents the hybrid systems based on PNI10@PNN (left), PMAM10@PNN (middle), and PNN10@PNN (right). Especially the last system shows a drastic change of the reaction rate. Reproduced from reference [68], American Chemical Society ACS Omega.

Similar to previous works, the obtained Arrhenius plots show that the swelling of the core and shell can influence the catalytic activity of the silver nanoparticles. As mentioned before, the cross-linker content of the core seems to be a very important parameter for the switchability of the catalytic activity. A higher cross-linker content of the core appears to be connected to a stronger influence of the carrier swelling/shrinking degree on the catalytic activity of the silver NPs.

Besides the often used reduction of *p*-nitrophenol also Congo red was used in a recent example of silver NPs in PS-core/MG-shell particles, where the NPs can be located only in the shell. Dyes are a major problem in regions with intense textile production due to release in the environment. Especially, Congo red is problematic since it is harmful for mammals. Hence, it would be desirable to have an efficient procedure to decompose this dye. The work by Naseem et al. reveals an enhanced reduction efficiency of Congo Red by this particular core-shell microgel dispersion type [104].

## 5. Core-shell microgels at surfaces and interfaces

In the previous sections we have already discussed structure and properties of core-shell microgels and also their application in catalysis. In addition to these topics, core-shell microgels deposited on surfaces e.g. as actuators or at interfaces to make smart emulsions are of growing relevance. This is of course related to the growing interest in adsorbed microgels and smart Pickering emulsions in general [105]. In this context recent work was published by Yi Gong and co-workers [106]. These authors studied the interfacial rheology of core-shell microgels at the decane-water interface. The particles were p(NIPAM-co-AAc)@poly-2,2,2-trifluoroethyl methacrylate (PTFMA) core-shell systems and the authors measured the interfacial pressure in a Langmuir trough. The microgel monolayers at the oil-water interface were exposed to compression expansion cycles. The work shows that the elastic behavior of the monolayer is dominated by the properties of the shell.

Also Vasudevan and co-workers have studied core-shell microgels in this case with silica cores of variable diameter and p(NIPAM) shells of variable thickness with respect to their behavior at the oil water interface [107]. First the authors presented a theoretical model for the wetting behavior of this type of inorganic-organic core-shell microgel as a function of the specific geometrical and compositional features of the different particles. The model is in line with experimental findings showing that the occupied area per core-shell microgel is strongly dependent on both varied parameters, namely the core size and shell thickness. Below a critical shell thickness, the cores are found to be located in a position where they just touch the oil-water interface. Beyond a critical shell thickness, the core detaches from the interface. Compared to hard sphere particles at the oil-water interface such hard-soft hybrids exhibit a subtle interplay between position at the interface, interfacial dynamics, and might have interesting applications in interfacial rheology. This is in line with the observations by Yi Gong et al. [106].

Besides the behavior of core-shell microgels at liquid-liquid interfaces their behavior at solid surfaces has also been studied. This topic is of great relevance when applications as sensors or actuators are envisaged.

In a work on particles made of a p(NIPMAM)-core and a p(NNPAM)-shell we have shown that these core-shell microgels still exhibit a linear swelling behaviour even if confined to a solid interface (Si-wafer) [57]. In this work we have used ellipsometry to measure the film thickness as a function of temperature while the wafers were immersed in water. Such surface coatings might be suitable actuating units in e.g. etalons [108] giving rise to a linear piezo-like response. Related to this, it is of course interesting to study adsorbed core-shell microgels with respect to their mechanical properties and changes. According to Schulte el al., this can be done by nano-indentation experiments [109]. For sensing applications adsorbed core-shell microgels can be used as templates for making composite films for surface enhanced Raman scattering (SERS). This was shown by Sen Yan and co-workers using poly(styrene-co-*N*-isopropylacrylamide)@polyacrylic acid microgels which were deposited in an ordered array and subsequently sputtered with an Au film [110]. Such a process yields very nice highly ordered surface coatings. This is shown in Figure 9.

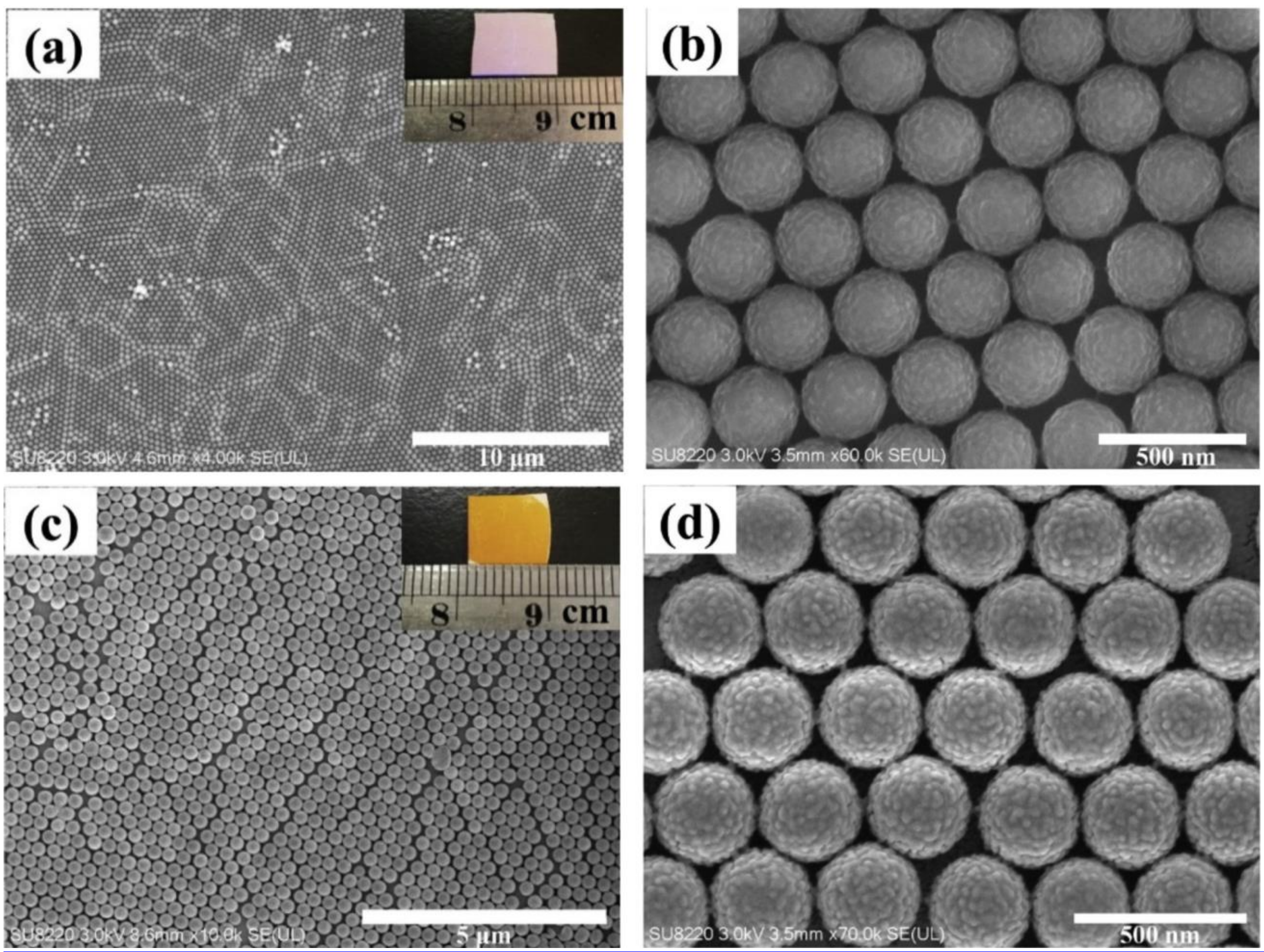

**Figure 9:** The upper row (images a, b) shows SEM micrographs of a PSN@PAA monolayered film. Images c, d show a PSN@PAA/Au composite film. The insets show the macroscopic optical appearance of the prepared films. The obtained hybrid film exhibits excellent SERS enhancement factors and are shown to be re-usable several times for sensing without any loss of performance. Reproduced with permission from reference [110]. Copyright 2020 Elsevier.

## 6. Conclusion and perspectives

On the occasion of this special issue in honor of the contributions of Matthias Ballauff and his group to the field of colloid and polymer science, we have organized this review on core-shell thermosensitive microgel particles taking contributions by Matthias and coworkers as starting point and guidance. We have deliberately focused our article on either soft copolymer microgel core-shell structures, or microgel shells grown onto hard nanoparticle cores, as recent progress using scattering techniques and in particular small-angle scattering with simulation-aided structural analysis coincides with our own research interests over the past years. Other topics which may be looked at as related, like protein or enzyme adsorption onto microgels also studied by Matthias Ballauff [16, 111, 112] have not been included in this review for lack of space, although other pioneering studies have been recently published by other groups active in this field [113, 50].

Due to our inclination towards scattering techniques – a trait of character we share with Matthias –, we have mostly restricted ourselves to core-shell particles of spherical symmetry. Naturally, nanorod structures as studied by Dong et al. [114], the groups of Dietsch and Schurtenberger [42, 43], and the ones of Ballauff and Müller [70] can also be apprehended in detail by scattering due to the high degree of symmetry. Computer simulations have certainly not been treated here at the level they deserve, as they now allow to follow the formation of complex topologies in detail [115, 116, 117, 118]. Finally, self-assembly of microgels, as well as their interfacial activity and their possible application as switchable cell culture substrates for vertebrate cells have also been omitted [119].

If one looks at recent progress in the field of microgels, one can only be surprised by the liveliness of the subject, several decades after their discovery, and at least twenty years of in-depth studies. Current directions of research seem to develop towards more complex topologies, like multi-shell or hollow structures with multi-responsiveness. Obviously, the main driving force comes from chemical synthesis, which can explore the infinite possibilities of organic structures. This will allow for fine-tuning of the interactions between microgel hosts and guest molecules, by introducing specific functions in well-defined positions inside the microgels. Scattering studies may then allow (or not, depending on contrast situations), to help designing and understanding the function of such structures. If hosting molecules or simply solvent has an impact on the intrinsic swelling properties of microgels, then one may also use them as sensor capable of detecting local concentration changes. Such studies may then be benchmarked by rapidly spreading high-resolution microscopy techniques, like dSTORM [80] [120], which however depend on the presence of fluorophores which might cause a stronger perturbation compared to simple deuteration. Whatever comes next, it looks like the field will continue its growth, much of the basis of which was laid by a small number of groups the work of which we have attempted to acknowledge and highlight here.

**Conflicts of interest**

There are no conflicts of interest to declare.

**Acknowledgements**

The authors are thankful for support by the joint ANR/DFG CoreShellGel project, Grant ANR-14-CE35-0008-01 of the French Agence Nationale de la Recherche, and Grant HE2995/5-1 by Deutsche Forschungsgemeinschaft.

**About the authors**

**Julian Oberdisse** is a soft matter physicist by training and he is research director in the soft matter group of Laboratoire Charles Coulomb (CNRS/University of Montpellier). Originally coming from surfactant science with his PhD with G. Porte in 1997, he started working on simulations of polymer systems during a postdoctoral stay in Naples with G. Marrucci. Nowadays, he studies the structure and dynamics of various soft matter systems in bulk, in solution, and sometimes at interfaces. His primary research interest is the investigation of nanostructures using scattering experiments, in polymer nanocomposites and, more recently, microgels, combined with modelling approaches often based on computer simulations.

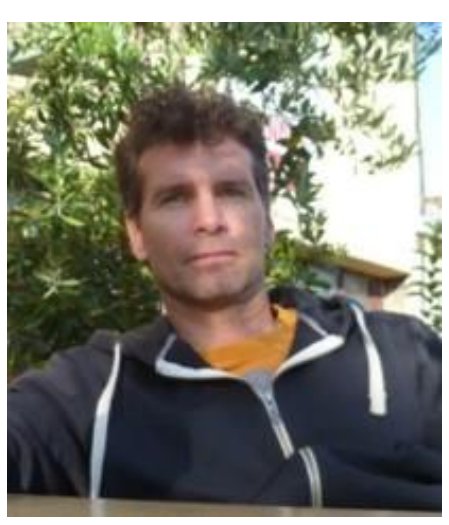

*Julian Oberdisse*

**Thomas Hellweg** is a physical chemist by training and received his PhD in 1995 at Bielefeld university in the group of T. Dorfmüller. From 1996 to 1998 he was a postdoctoral fellow at the Centre de Recherche Paul Pascal in Bordeaux in the groups of D. Langevin and D. Roux. In 1998 he started his habilitation at the TU Chemnitz in the physics department (in the Materials and liquids group; J.-B. Suck). In 2001 he moved to the Technical University of Berlin, where he finished his habilitation in 2005 in the Iwan-N.-Stranski-Lab. for Physical and Theoretical Chemistry in the group of G. Findenegg. In March 2007 he was appointed as professor for physical and colloid chemistry at the University of Bayreuth and in 2010 he moved back to Bielefeld university where he became full professor

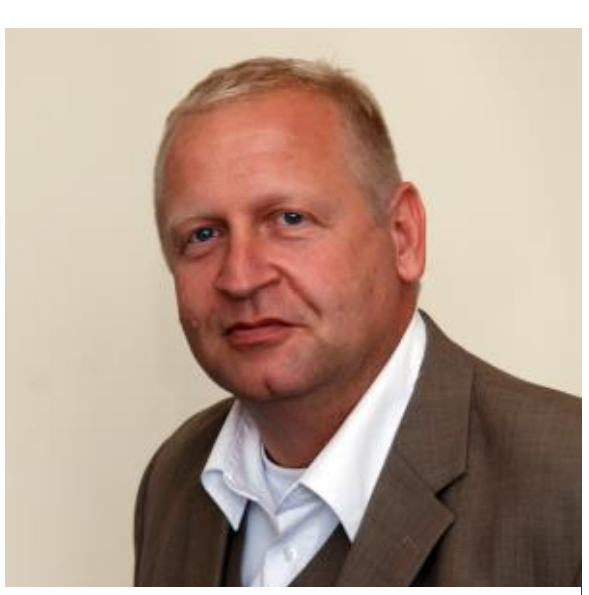

*Thomas Hellweg*

for physical and biophysical chemistry. His research is focusing on the physical properties of soft matter and he is an expert for scattering experiments with neutrons, X-rays, and light.

**References**

[1] R.H. Pelton, P. Chibante, Colloids and Surfaces **20**, 247 (1986).

[2] R. Pelton, Adv. Colloid Interf. Sci. **85**, 1 (2000).

[3] S. Nayak, L.A. Lyon, Angew. Chem. Int. Ed. **44**, 7686 (2005).

[4] M. Karg, A. Pich, T. Hellweg, T. Hoare, L.A. Lyon, J.J. Crassous, D. Suzuki, R.A. Gumerov, S. Schneider, I.I. Potemkin, W. Richtering, Langmuir **35**(19), 6231 (2019). doi: 10.1021/acs.langmuir.8b04304

[5] Q.M. Zhang, D. Berg, S.M. Mugo, M.J. Serpe, Chem. Commun. **51**, 9726 (2015)

[6] A. Scotti, U. Gasser, E.S. Herman, J. Han, A. Menzel, L.A. Lyon, A. Fernandez-Nieves, Physical Review E **96**(3) (2017). doi: 10.1103/physreve.96.032609

[7] U. Gasser, A. Fernandez-Nieves, Physical Review E **81**(5) (2010). doi: 10.1103/physreve.81.052401

[8] L.A. Lyon, J. D. Debord, S.B. Debord, C.D. Jones, J.G. McGrath, M.J. Serpe, J. Phys. Chem. B **108**, 19099 (2004)

[9] G. Fundueanu, M. Constantin, S. Bucatariu, P. Ascenzi, Macromolecular Chemistry and Physics **217**(22), 2525 (2016). doi: 10.1002/macp.201600324

[10] K. Uhlig, T. Wegener, J. He, M. Zeiser, J. Bookhold, I. Dewald, N. Godino, M. Jaeger, T. Hellweg, A. Fery, C. Duschl, Biomacromolecules **17**(3), 1110 (2016). DOI: 10.1021/acs.biomac.5b01728

[11] T. Ngai, S.H. Behrens, H. Auweter, Chem. Commun. pp. 331 – 333 (2005)

[12] T. Ngai, H. Auweter, S.H. Behrens, Macromolecules **39**, 8171 (2006)

[13] S. Fujii, E.S. Read, B.P. Binks, S.P. Armes, Advanced Materials **17**(8), 1014 (2005). doi: 10.1002/adma.200401641

[14] K. Geisel, A.A. Rudov, I.I. Potemkin, W. Richtering, Langmuir **31**(48), 13145 (2015). doi: 10.1021/acs.langmuir.5b03530

[15] Y. Lu, M. Ballauff, Progr. Polym. Sci. **26**, 767 (2011).

[16] N. Welsch, A.L. Becker, J. Dzubiella, M. Ballauff, Soft Matter **8**(5), 1428 (2012)

[17] S. Uchiyama, Y. Matsumura, A.P. de Silva, K. Iwai, Anal. Chem. **76**, 1793 (2004)

[18] B. Wedel, M. Zeiser, T. Hellweg, Zeitschrift f. Phys. Chem. **227**, 00 (2012)

[19] Y. Hertle, T. Hellweg, J. Mater. Chem. B **1**, 5874 (2013)

[20] T. Hoare, R. Pelton, Macromolecules **37**, 2544 (2004)

[21] T. Hoare, R. Pelton, Langmuir **20**, 2123 (2004)

[22] T. Hoare, R. Pelton, Langmuir **22**, 7342 (2006)

[23] J.H. Kim, M. Ballauff, Colloid Polym. Sci. **277**, 1210 (1999)

[24] N. Dingenouts, C. Nordhausen, M. Ballauff, Macromolecules **31**, 8912 (1998)

[25] Y. Lu, Y. Mei, M. Ballauff, M. Drechsler, J. Phys. Chem. B **110**, 3930 (2006).

[26] Y. Lu, A. Wittemann, M. Ballauff, M. Drechsler, Macromolecular Rapid Communications **27**(14), 1137 (2006). doi: 10.1002/marc.200600190

[27] Y. Lu, Y. Mei, M. Drechsler, M. Ballauff, Angew. Chem. **118**, 827 (2006)

[28] M. Dulle, S. Jaber, S. Rosenfeldt, A. Radulescu, S. Förster, P. Mulvaney, M. Karg, Physical Chemistry Chemical Physics **17**(2), 1354 (2015). doi: 10.1039/c4cp04816d

[29] M. Mueller, M. Tebbe, D.V. Andreeva, M. Karg, R.A.A. Puebla, N.P. Perez, A. Fery, Langmuir **28**, 9168 (2012)

[30] I. Berndt, W. Richtering, Macromolecules **36**, 8780 (2003)

[31] I. Berndt, J.S. Pedersen, P. Lindner, W. Richtering, Langmuir **22**, 459 (2006)

[32] T. Hellweg, Journal of Polymer Science Part B: Polymer Physics **51**(14), 1073 (2013). doi: 10.1002/polb.23294

[33] D. Suzuki, K. Horigome, T. Kureha, S. Matsui, T. Watanabe, Polymer Journal **49**(10), 695 (2017). doi: 10.1038/pj.2017.39

[34] M. Ballauff, Macromol. Chem. Phys. **204**, 220 (2003)

[35] J.J. Crassous, M. Ballauff, M. Drechsler, J. Schmidt, Y. Talmon, Langmuir **22**(6), 2403 (2006)

[36] D. Duracher, F. Sauzedde, A. Elaissari, A. Perrin, C. Pichot, Coll. Polym. Sci. **276**, 219 (1998).

[37] T. Hellweg, C.D. Dewhurst, W. Eimer, K. Kratz, Langmuir **20**(11), 4330 (2004)

[38] W. tao Hu, H. Yang, H. Cheng, H. qing Hu, Chinese Journal of Polymer Science **35**(9), 1156 (2017). doi: 10.1007/s10118-017-1969-7

[39] M. Karg, I. Pastoriza-Santos, L.M. Liz-Marzan, T. Hellweg, Chem. Phys. Chem. **7**, 2298 (2006)

[40] N. Nun, S. Hinrichs, M.A. Schroer, D. Sheyfer, G. Grübel, B. Fischer, Gels **3**(3), 34 (2017). doi: 10.3390/gels3030034

[41] L. Chen, L. Li, W. Liu, Y. Yang, X. Liu, Journal of Materials Research **29**(10), 1153 (2014). doi: 10.1557/jmr.2014.92

[42] C. Dagallier, H. Dietsch, P. Schurtenberger, F. Scheffold, Soft Matter **6**, 2174 (2010)

[43] V. Städele, U. Gasser, H. Dietsch, Soft Matter **8**(16), 4427 (2012). doi: 10.1039/c2sm07152e

[44] S. Louguet, B. Rousseau, R. Epherre, N. Guidolin, G. Goglio, S. Mornet, E. Duguet, S. Lecommandoux, C. Schatz, Polymer Chemistry **3**(6), 1408 (2012). doi: 10.1039/c2py20089a

[45] S. Wu, J. Kaiser, M. Drechsler, M. Ballauff, Y. Lu, Colloid and Polymer Science **291**(1), 231 (2012). doi: 10.1007/s00396-012-2736-5

[46] S. Wu, J. Dzubiella, J. Kaiser, M. Drechsler, X. Guo, M. Ballauff, Y. Lu, Angewandte Chemie International Edition **51**(9), 2229 (2012). doi: 10.1002/anie.201106515

[47] S. Nayak, D. Gan, M.J. Serpe, L.A. Lyon, Small **1**(4), 416 (2005)

[48] J. Dubbert, T. Honold, J.S. Pedersen, A. Radulescu, M. Drechsler, M. Karg, W. Richtering, Macromolecules **47**(24), 8700 (2014). doi: 10.1021/ma502056y

[49] J. Dubbert, K. Nothdurft, M. Karg, W. Richtering, Macromolecular Rapid Communications **36**(2), 159 (2015). doi: 10.1002/marc.201400495

[50] A.J. Schmid, J. Dubbert, A.A. Rudov, J.S. Pedersen, P. Lindner, M. Karg, I.I. Potemkin, W. Richtering, Scientific Reports **6**(1) (2016). doi: 10.1038/srep22736

[51] A. Scotti, M. Brugnoni, A.A. Rudov, J.E. Houston, I.I. Potemkin, W. Richtering, The Journal of Chemical Physics **148**(17), 174903 (2018). doi: 10.1063/1.5026100

[52] C.D. Jones, L.A. Lyon, Macromolecules **33**, 8301 (2000)

[53] D. Gan, L.A. Lyon, J. Am. Chem. Soc. **123**, 7511 (2001)

[54] C.D. Jones, L.A. Lyon, Macromolecules **36**(6), 1988 (2003). doi: 10.1021/ma021079q

[55] M. Zeiser, I. Freudensprung, T. Hellweg, Polymer **53**, 6096 (2012)

[56] S.M. Lee, Y.C. Bae, Macromolecules **47**(23), 8394 (2014). doi: 10.1021/ma5020897

[57] M. Cors, O. Wrede, A.C. Genix, D. Anselmetti, J. Oberdisse, T. Hellweg, Langmuir **33**(27), 6804 (2017). doi: 10.1021/acs.langmuir.7b01199

[58] I. Berndt, J.S. Pedersen, W. Richtering, Journal of the American Chemical Society **127**(26), 9372 (2005). doi: 10.1021/ja051825h

[59] M. Cors, L. Wiehemeier, Y. Hertle, A. Feoktystov, F. Cousin, T. Hellweg, J. Oberdisse, Langmuir **34**(50), 15403 (2018). doi: 10.1021/acs.langmuir.8b03217

[60] M. Cors, L. Wiehemeier, J. Oberdisse, T. Hellweg, Polymers **11**(4), 620 (2019). doi: 10.3390/polym11040620

[61] M. Cors, O. Wrede, L. Wiehemeier, A. Feoktystov, F. Cousin, T. Hellweg, J. Oberdisse, Scientific Reports **9**(1) (2019). doi: 10.1038/s41598-019-50164-6

[62] Y. Mei, Y. Lu, F. Polzer, M. Ballauff, M. Drechsler, Chem. Mater. **19**, 1062 (2007)

[63] Y. Lu, S. Proch, M. Schrinner, M. Drechsler, R. Kempe, M. Ballauff, J. Mater. Chem. **19**(23), 3955 (2009)

[64] G. Sharama, Y. Mei, Y. Lu, M. Ballauff, T. Irrgang, S. Proch, R. Kempe, Journal of Catalysis **246**(1), 10 (2007). doi: 10.1016/j.jcat.2006.11.016

[65] A. Pfaff, V.S. Shinde, Y. Lu, A. Wittemann, M. Ballauff, A.H.E. MÃ¼ller, Macromolecular Bioscience **11**(2), 199 (2010). doi: 10.1002/mabi.201000324

[66] D. Suzuki, H. Kawaguchi, Langmuir **21**(18), 8175 (2005). doi: 10.1021/la0504356

[67] D. Suzuki, H. Kawaguchi, Langmuir **21**(25), 12016 (2005). doi: 10.1021/la0516882

[68] T. BrÃ¤ndel, V. Sabadasch, Y. Hannappel, T. Hellweg, ACS Omega **4**(3), 4636 (2019). doi: 10.1021/acsomega.8b03511

[69] D. Suzuki, H. Kawaguchi, Colloid and Polymer Science **284**(12), 1443 (2006). doi: 10.1007/s00396-006-1523-6

[70] Y. Xu, J. Yuan, B. Fang, M. Drechsler, M. Müllner, S. Bolisetty, M. Ballauff, A.H.E. Müller, Advanced Functional Materials **20**(23), 4182 (2010). doi: 10.1002/adfm.201000769

[71] S. Thies, P. Simon, I. Zelenina, L. Mertens, A. Pich, Small **14**(51), 1803589 (2018). doi: 10.1002/smll.201803589

[72] D. Suzuki, C. Kobayashi, Langmuir **30**(24), 7085 (2014). doi: 10.1021/la5017752

[73] C. Kobayashi, T. Watanabe, K. Murata, T. Kureha, D. Suzuki, Langmuir **32**(6), 1429 (2016). doi: 10.1021/acs.langmuir.5b03698

[74] T. Watanabe, C. Song, K. Murata, T. Kureha, D. Suzuki, Langmuir **34**(29), 8571 (2018). doi: 10.1021/acs.langmuir.8b01047

[75] X. Wu, R.H. Pelton, A.E. Hamielec, D.R. Woods, W. McPhee, Colloid Polym. Sci. **272**, 467 (1994)

[76] S. Meyer, W. Richtering, Macromolecules **38**, 1517 (2005)

[77] L. Wiehemeier, M. Cors, O. Wrede, J. Oberdisse, T. Hellweg, T. Kottke, Phys. Chem. Chme. Phys. **21**, 572 (2019). DOI: 10.1039/C8CP05911J

[78] S. Bergmann, O. Wrede, T. Huser, T. Hellweg, Phys. Chem. Chem. Phys. **20**, 5074 (2018). Doi:10.1039/C7CP07648G

[79] S. Matsui, Y. Nishizawa, T. Uchihashi, D. Suzuki, ACS Omega **3**(9), 10836 (2018). doi: 10.1021/acsomega.8b01770

[80] A.P.H. Gelissen, A. Oppermann, T. Caumanns, P. Hebbeker, S.K. Turnhoff, R. Tiwari, S. Eisold, U. Simon, Y. Lu, J. Mayer, W. Richtering, A. Walther, D. Wöll, Nanoletters **16**, 7295 (2016). 10.1021/acs.nanolett.6b03940

[81] A.M. Hecht, R. Duplessix, E. Geissler, Macromolecules **18**, 2167 (1985)

[82] I. Berndt, J.S. Pedersen, W. Richtering, Angew. Chem. **118**, 1769 (2006)

[83] I. Berndt, C. Popescu, F.J. Wortmann, W. Richtering, Angewandte Chemie International Edition **45**(7), 1081 (2006). doi: 10.1002/anie.200502893

[84] N. Boon, P. Schurtenberger, Phys. Chem. Chem. Phys. **19**, 23740 (2017). DOI: 10.1039/c7cp02434g

[85] R. Keidel, A. Ghavami, D.M. Lugo, G. Lotze, O. Virtanen, P. Beumers, J.S. Pedersen, A. Bardow, R.G. Winkler, W. Richtering, Science Advances **4**(4), eaao7086 (2018). doi: 10.1126/sciadv.aao7086

[86] O.L.J. Virtanen, A. Mourran, P.T. Pinard, W. Richtering, Soft Matter **12**(17), 3919 (2016). doi: 10.1039/c6sm00140h

[87] L. Lienafa, J. Oberdisse, S. Mora, S. Monge, J.J. Robin, Macromolecules **44**(13), 5326 (2011)

[88] B. Hammouda, Macromolecular Theory and Simulations **21**(6), 372 (2012). doi: 10.1002/mats.201100111

[89] M. Cors, L. Wiehemeier, O. Wrede, A. Feoktystov, F. Cousin, T. Hellweg, J. Oberdisse, Soft Matter **in press** (2020)

[90] A. Balaceanu, D.E. Demco, M. Möller, A. Pich, Macromol. Chem. Phys. **212**, 2467 (2011)

[91] Y. Wu, S. Wiese, A. Balaceanu, W. Richtering, A. Pich, Langmuir **30**(26), 7660 (2014). doi: 10.1021/la501181k

[92] J.J. Crassous, A.M. Mihut, L.K. Mansson, P. Schurtenberger, Nanoscale **7**, 15971 (2015)

[93] B.R. Saunders, B. Vincent, Adv. Colloid Interf. Sci. **80**, 1 (1999)

[94] Y. Hirokawa, T. Tanaka, J. Chem. Phys. **81**(12), 6379 (1984)

[95] Y. Lu, M. Yu, M. Drechsler, M. Ballauff, Macromol. Symp. **254**, 97 (2007)

[96] N. Welsch, A. Wittemann, M. Ballauff, J. Phys. Chem. B **113**(49), 16039 (2009)

[97] T. Kureha, Y. Nagase, D. Suzuki, ACS Omega **3**(6), 6158 (2018). doi: 10.1021/acsomega.8b00819

[98] Y. Wang, L. Wang, J. Hao, S. Dong, New Journal of Chemistry **42**, 2149 (2018). doi: 10.1039/c7nj03661b

[99] M. Karg, S. Wellert, S. Prevost, R. Schweins, C. Dewhurst, L.M. Liz-Marzán, T. Hellweg, Coll. Polym. Sci. **289**(5), 699 (2011)

[100] M. Das, N. Sanson, D. Fava, E. Kumacheva, Langmuir **23**, 196 (2007).

[101] M. Das, L. Mardoukhovski, E. Kumacheva, Adv. Mater. **20**, 2371 (2008)

[102] L. L. Tzounis, M. Dona, J. Lopez-Romero, A. Fery, R. Contreras-Caceres, Applied Materials & Interfaces **11**, 29360 (2019). doi: 10.1021/acsami.9b10773

[103] L.Q. Yang, M.M. Hao, H.Y. Wang, Y. Zhang, Colloid and Polymer Science **293**(8), 2405 (2015). doi: 10.1007/s00396-015-3642-4

[104] K. Naseem, Z.H. Farooqi, R. Begum, W. Wu, A. Irfan, A.G. Al-Sehemi, Macromolecular Chemistry and Physics **219**(18), 1800211 (2018). doi: 10.1002/macp.201800211

[105] S. Wiese, Y. Tsvetkova, N.J. Daleiden, A.C. Spieß, W. Richtering, Colloids and Surfaces A: Physicochemical and Engineering Aspects **495**, 193 (2016). doi: 10.1016/j.colsurfa.2016.02.003

[106] Y. Gong, Z. Zhang, J. He, Industrial & Engineering Chemistry Research **56**(50), 14793 (2017). doi: 10.1021/acs.iecr.7b03963

[107] S.A. Vasudevan, A. Rauh, M. Kröger, M. Karg, L. Isa, Langmuir **34**(50), 15370 (2018). doi: 10.1021/acs.langmuir.8b03048

[108] Y. Zhang, W.S. Carvalho, C. Fang, M.J. Serpe, Sensors and Actuators B: Chemical **290**, 520 (2019). doi: 10.1016/j.snb.2019.03.147

[109] M.F. Schulte, A. Scotti, A.P.H. Gelissen, W. Richtering, A. Mourran, Langmuir **34**(14), 4150 (2018). doi: 10.1021/acs.langmuir.7b03811

[110] S. Yan, R. An, Y. Zou, N. Yang, Y. Zhang, Sensors and Actuators B: Chemical **302**, 127107 (2020). doi: 10.1016/j.snb.2019.127107

[111] N. Welsch, Y. Lu, J. Dzubiella, M. Ballauff, Polymer **54**(12), 2835 (2013). doi: 10.1016/j.polymer.2013.03.027

[112] S. Angioletti-Uberti, M. Ballauff, J. Dzubiella, Molecular Physics **116**(21-22), 3154 (2018). doi: 10.1080/00268976.2018.1467056

[113] S. Matsui, K. Hosho, H. Minato, T. Uchihashi, D. Suzuki, Chemical Communications **55**(68), 10064 (2019). doi: 10.1039/c9cc05116c

[114] X. Dong, X. Zou, X. Liu, P. Lu, J. Yang, D. Lin, L. Zhang, L. Zha, Colloids and Surfaces A: Physicochemical and Engineering Aspects **452**, 46 (2014). doi: 10.1016/j.colsurfa.2014.03.090

[115] A.J. Moreno, F.L. Verso, Soft Matter **14**(34), 7083 (2018). doi: 10.1039/c8sm01407h

[116] V.Y. Rudyak, A.A. Gavrilov, E.Y. Kozhunova, A.V. Chertovich, Soft Matter **14**(15), 2777 (2018). doi: 10.1039/c8sm00170g

[117] A.P.H. Gelissen, A. Scotti, S.K. Turnhoff, C. Janssen, A. Radulescu, A. Pich, A.A. Rudov, I.I. Potemkin, W. Richtering, Soft Matter **14**(21), 4287 (2018). doi: 10.1039/c8sm00397a

[118] L. Rovigatti, N. Gnan, L. Tavagnacco, A.J. Moreno, E. Zaccarelli, Soft Matter **15**(6), 1108 (2019). doi: 10.1039/c8sm02089b

[119] S. Schmidt, M. Zeiser, T. Hellweg, C. Duschl, A. Fery, H. Möhwald, Adv. Func. Mater. **20**(19), 3235 (2010)

[120] P. Otto, S. Bergmann, A. Sandmeyer, M. Dirksen, O. Wrede, T. Hellweg, T. Huser, Nanoscale Advances **2**(1), 323 (2020). doi: 10.1039/c9na00670b